\documentclass[final,5p,times,twocolumn]{elsarticle}
\usepackage{lineno,hyperref,amsmath,amsthm,amssymb,amsfonts,ragged2e,color,subfig}
\journal{``High Energy Density Physics"}
\begin{document}
\begin{frontmatter}
\title{Rogue waves in multi-pair plasma medium}
\author{S. Khondaker$^*$, A. Mannan,  N. A. Chowdhury, and A. A. Mamun}
\address{Department of Physics, Jahangirnagar University, Savar, Dhaka-1342, Bangladesh\\
Email: $^*$khondakershohana03@gmail.com}
\begin{abstract}
The nonlinear propagation of ion-acoustic (IA) waves (IAWs), which are governed by the nonlinear
Schr\"{o}dinger equation (NLSE), in multi-pair plasmas (MPPs) containing adiabatic positive and negative  ion fluids as well as
non-extensive ($q$-distributed) electrons and positrons, is theoretically investigated.
It is observed that the MPP under consideration supports two types of modes (namely, fast and
slow IA modes), and the modulationally stable and unstable parametric regimes for the fast and slow IA modes are determined by
the sign of the ratio of the dispersive coefficient to the nonlinear one. It is also found
that the modulationally unstable regime generates highly energetic IA rogue waves (IARWs),
and the amplitude as well as the width of the IARWs decrease with increase in the value of  $q$ (for both $q>0$ and $q<0$ limits).
These new striking features of the IARWs are found to be applicable in the space [viz.
D-region ($\rm H^+, O_2^-$) and F-region ($\rm H^+, H^-$) of the Earth's ionosphere] and
laboratory MPPs [viz. fullerene ($\rm C^+, C^-$)].
\end{abstract}
\end{frontmatter}
\section{Introduction}
\label{1sec:int}
Recently, the painstaking observational results support the existence of the multi-pair plasmas (MPPs)
not only in space environments (viz. cometary comae \cite{Chaizy1991}, chromosphere, upper regions of Titan's atmosphere \cite{El-Labany2012},
solar wind, D-region ($\rm H^+, O_2^-$) and F-region ($\rm H^+, H^-$) of the Earth's ionosphere \cite{Elwakil2010}, etc.)
but also laboratory situations (viz. Fullerene ($\rm C^+, C^-$) \cite{Sabry2008}, plasma processing
reactors \cite{Gottscho1986}, neutral beam sources \cite{Bacal1979}, etc.). Plasma physicists are solely
practicing wave dynamics, basically, ion-acoustic (IA) waves (IAWs) in which restoring force is
provided by the electrons thermal pressure and moment of inertia is provided by the pair-ion (PI) mass density. The
adiabatic presure/dissipative force from the inertial component of the plasma system provides a great effect
to generate various nonlinear electrostatic structures (viz. shocks, solitons, and envelope solitons) in
MPPs. The existence of positive ions in MPPs drastically alters charge neutrality condition \cite{Chowdhury2017},
the dispersive property  \cite{El-Labany2012}, and nonlinearity of the IAWs  \cite{Chowdhury2017}.

Boltzmann-Gibbs (BG) statistics, which was successful in describing thermally equilibrium system,
did not describe many natural and artificial complex systems that are far from equilibrium.
Motivated by this situation, Tsalli came up with a concept of generalized BG statistics  known as
non-extensive  statistical mechanics that has offered a theoretical basis for interpreting and
analyzing non-equilibrium complex systems \cite{Tsallis1988}. In non-extensive $q$-distribution, the non-extensivity ($q$)
plays an indispensable role in executing the long range inter particle forces which includes
Newtonian gravitational forces and coulomb electric forces among charged particles in a plasma medium.
Many researchers have given more attention on $q$-distribution to study the nonlinear property
of the plasma medium \cite{Jannat2016,Tribeche2010,Bains2011}. Jannat \textit{et al.} \cite{Jannat2016} studied
Gardner Solitons (GSs) in a non-extensive PI plasma (PIP).  Tribeche \textit{et al.} \cite{Tribeche2010} reported IA
solitary structures in two component plasma in presence of $q$-distributed electrons, and observed that
the amplitude of the potential pulse increases and more spiky with an increase
in $q$. Ghosh \textit{et al.} \cite{Ghosh2013} studied the effects of non-extensivity on the
GSs in a non-extensive electrons and positrons plasma medium.
Eslami \textit{et al.} \cite{Eslami2011} investigated the stability of the IAWs
in three component plasmas with $q$-distributed electrons and positrons.

The field of modulational instability (MI) and rogue waves (RWs) has considered one of the
most interdisciplinary areas of research enclosing plasma physics \cite{Chowdhury2017}, oceanography \cite{Kharif2009},
Bose-Einstein condensation \cite{Bludov2009}, optics \cite{Solli2007}, superfluid helium \cite{Ganshin2008}, and
even finance \cite{Yan2010} . Researchers have devoted their attention to solve the mystery of this
colossal waves due to the intrinsically arbitrary nature, and the complex formation mechanisms. Recently,
a number of authors have studied different criteria of MI and RWs  in space and laboratory plasmas.
Bains \textit{et al.} \cite{Bains2011} studied the MI of IAWs with a $q$-distributed electrons,
and found that the stable domain for IAWs  increases with $q$ within sub-extensive limit.
Abdelwahed \textit{et al.} \cite{Abdelwahed2016} studied IA RWs (IARWs) in PIP in presence of super-thermal
electrons. Elwakil \textit{et al.} \cite{Elwakil2010} analyzed the effects of non-thermal electrons
on  MI conditions of IAWs, and observed that electrons non-thermality reduces the critical wave
number ($k_c$) in PIP. El-Labany \textit{et al.} \cite{El-Labany2012} have derived the nonlinear
Schr\"{o}dinger equation (NLSE) in a three component PIP in presence of iso-thermal electrons, and
observed that the number density of the negative ions enhances the amplitude of the IARWs. In our
present investigation, we have extended the previous work of El-Labany \textit{et al.} \cite{El-Labany2012}
by considering adiabatic pressure of the PI in four component (comprising $q$-distributed  electrons and positrons, adiabatic
positive and negative ions) PIP, and also will examine the effects of the plasma parameters on the  MI of IAWs,
and the formation of the IARWS.

The outline of the paper is as follows: The governing equations describing our plasma
model are presented in Section \ref{2sec:Governing Equations}.  MI and RWs
are given in Section \ref{2sec:MI analysis and Rogue waves}. A brief discussion is
finally provided in Section \ref{2sec:Discussion}.
\section{Governing Equations}
\label{2sec:Governing Equations}
We consider an unmagnetized,  one-dimensional collisionless four component MPPs medium comprising
of inertialess $q$-distributed electrons (mass $m_e$; charge $q_e=-e$) and positrons
(mass $m_p$; charge $q_p=+e$), inertial adiabatic negative ions (mass $m_-$; charge
$q_-=-eZ_-$) as well as positive ions (mass $m_+$; charge $q_+=+eZ_+$), respectively. Here, $Z_-$ ($Z_+$) is the charge
state of negative (positive) ions. At equilibrium, the charge neutrality yields the condition $ Z_+ n_{+0}+n_{p0} = Z_- n_{-0}+ n_{e0}$;
where $n_{+0}$, $n_{-0}$, $ n_{e0}$, and $n_{p0}$, are the equilibrium number densities of positive ions,
negative ions, and $q$-distributed electrons and positrons, respectively. Now, the basic set of equations
in the normalized form can be written as
\begin{eqnarray}
&&\hspace*{-1.3cm}\frac{\partial n_{-}}{\partial t}+\frac{\partial}{\partial x}(n_{-} u_{-})=0,
\label{2eq:1}\\
&&\hspace*{-1.3cm}\frac{\partial u_{-}}{\partial t}+ u_{-}\frac{\partial u_{-} }{\partial x}+3 \sigma_1 n_{-} \frac{\partial n_{-}}{\partial x}= \frac{\partial \phi}{\partial x},
\label{2eq:2}\\
&&\hspace*{-1.3cm}\frac{\partial n_{+}}{\partial t}+\frac{\partial}{\partial x}(n_{+} u_{+})=0,
\label{2eq:3}\\
&&\hspace*{-1.3cm}\frac{\partial u_{+}}{\partial t}+ u_{+}\frac{\partial u_{+} }{\partial x}+3 \sigma_2 n_{+} \frac{\partial n_{+}}{\partial x}= - \mu_1 \frac{\partial \phi}{\partial x},
\label{2eq:4}\\
&&\hspace*{-1.3cm}\frac{\partial^2 \phi}{\partial x^2}= (\mu_2+\mu_p-1) n_e -\mu_pn_p+ n_{-} - \mu_2 n_{+}.
\label{2eq:5}
\end{eqnarray}
\begin{figure}[!tbp]
  \centering
  \subfloat[]{\includegraphics[width=60mm]{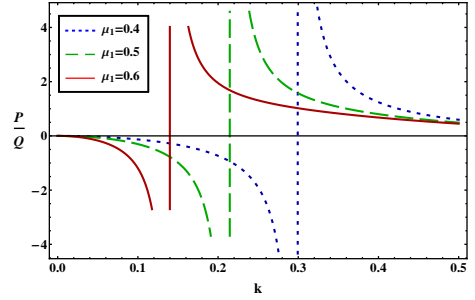}
  \label{2Fig:F1}}
  \hfill
  \subfloat[]{\includegraphics[width=60mm]{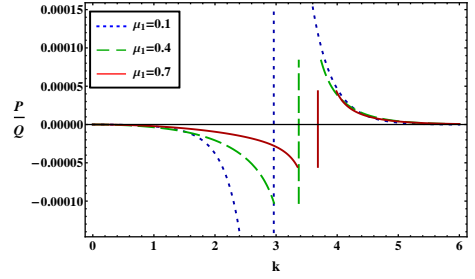}
  \label{2Fig:F2}}
  \caption{The variation of $P/Q$ with $k$ at different values of $\mu_1$ for (a) fast mode; (b) slow mode;
  along with fixed values of ~~$\delta=1.3$, $\mu_2=1.2$, $\mu_p=0.3$, $\sigma_1=0.005$, $\sigma_2=0.04$, and $q=1.5$.}
\end{figure}
Here, the normalizing and associated parameters are defined as: $n_-=N_-/n_{-0}$, $n_+=N_+/n_{+0}$,
$n_e=N_e/n_{e0}$, $n_p=N_p/n_{p0}$, $u_-=U_-/C_-$, $u_+=U_+/C_-$,
$\phi=e \tilde{\Phi}/k_BT_e$, $x=X/\lambda_{D_-}$, $t=T\omega_{P_-}$, $C_-=\sqrt{(Z_-k_BT_e/m_-)}$,
$\lambda_{D_-} = \sqrt{k_BT_e/4\pi e^2Z_-n_{-0}}$, $\omega_{P_-}=\sqrt{(4\pi e^2Z_-^2n_{-0}/{m_-})}$,
$P_-=P_{-0}\left(N_-/n_{-0}\right)^{\gamma}$, $P_+=P_{+0}\left(N_+/n_{+0}\right)^{\gamma}$, $P_{-0}=n_{-0}k_BT_-$,
$P_{+0}=n_{+0}k_BT_+$, $\gamma=(N+2)/N$, $\mu_1=Z_+m_-/Z_-m_+$, $\mu_2=Z_+n_{+0}/Z_-n_{-0}$,
$\mu_p=n_{p0}/Z_-n_{-0}$, $\sigma_1=T_-/Z_-T_e$, $\sigma_2=T_+m_-/Z_-T_e m_+$;
where $n_-$, $n_+$, $u_-$, $u_+$, $x$, $t$, $\phi$, $C_-$, $\lambda_{D-}$,
$\omega_{P-}$, $k_B$ $T_-$, $T_+$, $T_e$, $T_p$, $P_{-0}$, and $P_{+0}$  is the
number densities of negative ion, positive ion, electron, positron, negative ions fluid speed, positive ions
fluid speed, space co-ordinate, time co-ordinate, electro-static potential, sound speed
of the negative ions,  Debye length of the negative ions, angular frequency
of the negative ions, Boltzmann constant, negative ions temperature, positive ions temperature,
$q$-distributed electron temperature, $q$-distributed positron temperature, the equilibrium adiabatic
pressure of negative ions, and  the equilibrium adiabatic
pressure of positive ions, respectively. $N$ is the degree of freedom chosen to be $1$ for one-dimensional adiabatic case,
hence $\gamma$=3. It may be pointed here that the mass of the positive ion is greater than the negative ion,
i.e., $m_+>m_-$, the number density of the positive ions is greater than the negative ions, i.e., $n_{+0}>n_{-0}$,
and electron and positron temperatures are greater than the negative and positive ion temperature,
i.e., $T_e$, $T_p > T_-$, $T_+$. Now, the expressions for the normalized electron and positron number
density obeying $q$-distribution is given by \cite{Chowdhury2017}
\begin{eqnarray}
&&\hspace*{-1.3cm}n_e=\left[1 + (q-1)\phi \right]^{\frac{(q+1)}{2(q-1)}}=1+ n_1 \phi+ n_2\phi^2 +n_3 \phi^3+ \cdot\cdot\cdot,\\
&&\hspace*{-1.3cm}n_p=\left[1 - (q-1)\delta\phi \right]^{\frac{(q+1)}{2(q-1)}}=1- n_1 \delta\phi+ n_2\delta^2\phi^2 -n_3 \delta^3\phi^3+ \cdot\cdot\cdot,
\label{2eq:6}
\end{eqnarray}
where
\begin{eqnarray}
&&\hspace*{-1.3cm}n_1 = (1+q)/2,~~~n_2 = - (1+q)(q-3)/8,
\nonumber\\
&&\hspace*{-1.3cm}n_3 = (1+q)(q-3)(3q-5)/48,~~~\delta=T_e/T_p,~~~\mbox{and}~~~T_e>T_p.
\nonumber\
\end{eqnarray}
Here, the parameter $q$, generally known as entropic index, quantifies the degree of non-extensivity.
It is noteworthy that when $q=1$, the entropy reduces to standard Maxwell-Boltzmann
distribution. On the other hand, in the limits $q>1$ ($q<1$), the entropy shows sub-extensivity (super-extensivity).
Now, by substituting  \eqref{2eq:6} into \eqref{2eq:5} and expanding up to third order, we get
\begin{eqnarray}
&&\hspace*{-1.3cm}\frac{\partial^2 \phi}{\partial x^2} = (\mu_2-1)+  n_{-}-\mu_2 n_{+} + \gamma_1\phi+ \gamma_2 \phi^2+ \gamma_3 \phi^3\cdot\cdot\cdot,
\label{2eq:7}
\end{eqnarray}
where
\begin{eqnarray}
&&\hspace*{-1.3cm}\gamma_1=n_1(\mu_2+\mu_p-1+\mu_p\delta),
\nonumber\\
&&\hspace*{-1.3cm}\gamma_2= n_2(\mu_2+\mu_p-1-\mu_p\delta^2),
\nonumber\\
&&\hspace*{-1.3cm}\gamma_3 = n_3(\mu_2+\mu_p-1+\mu_p\delta^3).
\nonumber\
\end{eqnarray}
To analyze the MI of the IAWs, we will derive the NLSE by employing the reductive perturbation method.
So, the independent stretched variables as $\xi=\epsilon(x-v_gt)$ and $\tau=\epsilon^2t$,
where $\epsilon$ is a small perturbation parameter and $v_g$ is the envelope group velocity.
Furthermore, the dependent variables can be expanded as
\begin{eqnarray}
&&\hspace*{-1.3cm}\Lambda(x,t)=\Lambda_0+ \sum_{m=1}^{\infty}\epsilon^{(m)} \sum_{l=-\infty}^{\infty} \Lambda_{l}^{m}(\xi,\tau) \exp(il\Upsilon)
\label{2eq:8}
\end{eqnarray}
where $\Lambda_{l}^{m}$ = $[n_{-l}^{(m)}, u_{-l}^{(m)}, n_{+l}^{(m)}, u_{+l}^{(m)}, \phi_{l}^{(m)}]^T$, $\Lambda_0$ = $[1,0,1,0,0]^T$,
$\Upsilon=(kx-\omega t)$, and $\omega$ ($k$) represents the angular frequency
(wave number) of the carrier waves, respectively. We are going to parallel mathematical
steps as Chowdhury \textit{et al.} \cite{Chowdhury2018} have done in their work to find successively
the IAWs dispersion relation, group velocity, and NLSE. The IAWs dispersion relation
\begin{eqnarray}
&&\hspace*{-1.3cm} \omega^2=\frac{M \pm k^2\sqrt{M^2-4GN}}{2G},
\label{2eq:9}
\end{eqnarray}
where $G=(\gamma_1+k^2)$, $M=(1+\mu_1\mu_2+\gamma_1\lambda_1+\gamma_1\lambda_2+\lambda_1k^2+\lambda_2k^2)$,
$N=(\lambda_2+\mu_1\mu_2\lambda_1+\gamma_1\lambda_1\lambda_2+\lambda_1\lambda_2k^2)$, $\lambda_1=3\sigma_1$, and $\lambda_2=3\sigma_2$.
In order to get real and positive value of $\omega$, the condition $M^2>4GN$ is required to maintain.
It is clear from \eqref{2eq:9}, there exist two distinct modes  that depend on the signs $\pm$. The positive sign refers to fast
($\omega_f$) IA mode whereas negative sign refers to slow ($\omega_s$) IA mode. The group velocity of the IAWs is given by
\begin{eqnarray}
&&\hspace*{-1.3cm}v_g=\frac{\partial \omega}{\partial k}=\frac{F_1-2A^2S^2-AS(A-\mu_1\mu_2 S)}{2\omega k(A^2+\mu_1\mu_2S^2)},
\label{2eq:10}
\end{eqnarray}
Where $F_1=\omega^2(A^2+\mu_1\mu_2S^2)+k^2(\lambda_1A^2+\mu_1\mu_2\lambda_2S^2)$, $S = \lambda_1 k^2 - \omega^2$,
and $A = \omega^2 - \lambda_2 k^2$.
Finally, we obtain the standard NLSE:
\begin{eqnarray}
&&\hspace*{-1.3cm}i\frac{\partial \Phi}{\partial \tau}+P\frac{\partial^2\Phi}{\partial \xi^2}+Q|\Phi|^2\Phi=0,
\label{2eq:11}
\end{eqnarray}
where $\Phi=\phi_1^{(1)}$ for simplicity. The dispersion coefficient $P$ and the nonlinear coefficient $Q$ are, respectively, given by
\begin{eqnarray}
&&\hspace*{-1.3cm}P=\frac{F_2-A^3S^3}{2AS\omega k^2(A^2+\mu_1\mu_2S^2)},
\label{2eq:12}
\end{eqnarray}
\begin{eqnarray}
&&\hspace*{-1.3cm}Q=\frac{A^2S^2\{2 \gamma_2 (C_5+C_{10})+3\gamma_3-F_3\}}{2\omega k^2(A^2+\mu_1\mu_2S^2)},
\label{2eq:13}
\end{eqnarray}
where
\begin{eqnarray}
&&\hspace*{-1.3cm} F_2=A^3\{(\omega v_g-\lambda_1k)(\lambda_1k^3+k\omega^2-2\omega v_gk^2-kS)
\nonumber\\
&&\hspace*{-0.5cm} +(v_gk-\omega)(\omega^3+\lambda_1\omega k^2-2v_gk\omega^2-v_gkS)\}
\nonumber\\
&&\hspace*{-0.5cm} -\mu_1\mu_2S^3\{(\omega v_g-\lambda_2k)(\lambda_2k^3+k\omega^2-2\omega v_gk^2+kA)
\nonumber\\
&&\hspace*{-0.5cm} +(v_gk-\omega)(\omega^3+\lambda_2\omega k^2-2v_gk\omega^2+kv_gA)\},
\nonumber\\
&&\hspace*{-1.3cm} F_3=\frac{2\omega k^3(C_2+C_7)}{S^2}+\frac{(\omega^2 k^2+\lambda_1k^4)(C_1+C_6)}{S^2}
\nonumber\\
&&\hspace*{-0.5cm} +\frac{\mu_1\mu_2(\omega^2 k^2+\lambda_2k^4)(C_3+C_8)}{A^2}
\nonumber\\
&&\hspace*{-0.5cm} +\frac{2\mu_1\mu_2\omega k^3(C_4+C_9)}{A^2},
\nonumber\
\end{eqnarray}
\begin{eqnarray}
&&\hspace*{-1.3cm}C_1=\frac{2C_5k^2S^2-3\omega^2k^4-\lambda_1 k^6}{2 S^3},
\nonumber\\
&&\hspace*{-1.3cm}C_2=\frac{\omega(C_1S^2-k^4)}{k S^2},
\nonumber\\
&&\hspace*{-1.3cm}C_3=\frac{2C_5\mu_1k^2 A^2+\mu_1^2(\lambda_2k^6+3\omega^2k^4)}{2A^3},
\nonumber\\
&&\hspace*{-1.3cm}C_4=\frac{\omega(C_3A^2-\mu_1^2k^4)}{kA^2},
\nonumber\\
&&\hspace*{-1.3cm}C_5=\frac{A^3(3\omega^2k^4+\lambda_1k^6-2\gamma_2S^3)+\mu_2S^3(3\omega^2\mu_1^2k^4+\lambda_2\mu_1^2k^6)}{2A^3S^3(4k^2+\gamma_1)+2k^2S^2A^3-\mu_1\mu_2k^2A^2S^3},
\nonumber\\
&&\hspace*{-1.3cm}C_6=\frac{k^2\omega^2+2v_g\omega k^3+\lambda_1k^4-C_{10}S^2}{S^2(v_g^2-\lambda_1)},
\nonumber\\
&&\hspace*{-1.3cm}C_7=\frac{v_gC_6S^2-2\omega k^3}{S^2},
\nonumber\\
&&\hspace*{-1.3cm}C_8=\frac{\mu_1^2(2\omega v_g k^3+k^2\omega^2+\lambda_2k^4)-\mu_1C_{10}A^2}{A^2(v_g^2-\lambda_2)},
\nonumber\\
&&\hspace*{-1.3cm}C_9=\frac{v_gC_8A^2-2\omega \mu_1^2k^3}{A^2},
\nonumber\\
&&\hspace*{-1.3cm}C_{10}=\frac{F_4+2\gamma_2A^2S^2(v_g^2-\lambda_1)(v_g^2-\lambda_2)}
{F_5-\gamma_1A^2S^2(v_g^2-\lambda_1)(v_g^2-\lambda_2)},
\nonumber\\
&&\hspace*{-1.3cm} F_4=A^2(k^2\omega^2+ 2v_g\omega k^3+\lambda_1k^4)(v_g^2-\lambda_2)
\nonumber\\
&&\hspace*{-0.6cm} -\mu_2S^2(\mu_1^2k^2\omega^2+2\omega v_g \mu_1^2k^3+\lambda_2\mu_1^2k^4)(v_g^2-\lambda_1),
\nonumber\\
&&\hspace*{-1.3cm}F_5=A^2S^2(v_g^2-\lambda_2)+\mu_1\mu_2A^2S^2(v_g^2-\lambda_1).
\nonumber
\end{eqnarray}
\begin{figure}[!tbp]
  \centering
  \subfloat[]{\includegraphics[width=60mm]{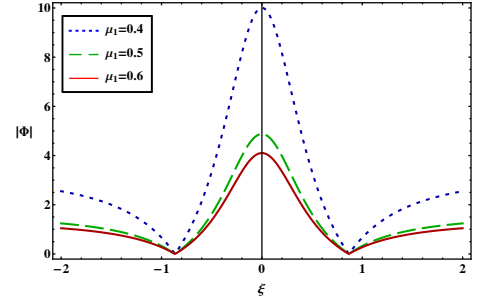}
  \label{2Fig:F3}}
  \hfill
  \subfloat[]{\includegraphics[width=60mm]{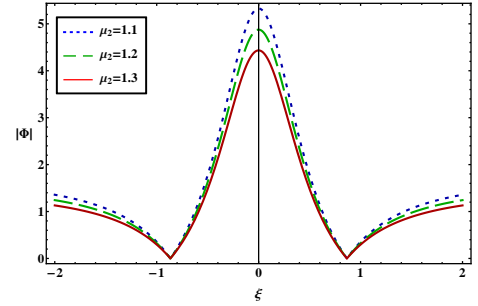}
  \label{2Fig:F4}}
   \caption{ The variation of $|\Phi|$ with $\xi$ at different values of (a) $\mu_1$ and $\mu_2=1.2$; (b) $\mu_2$ and $\mu_1=0.5$;
  along with $\delta=1.3$, $\mu_p=0.3$, $\sigma_1=0.005$, $\sigma_2=0.04$, $q=1.5$, $k=0.35$, $\tau=0$, and $\omega_f$.}
\end{figure}
\section{MI analysis and Rogue waves}
\label{2sec:MI analysis and Rogue waves}
The stability of IAWs in four component plasma medium is governed by the sign of $P$
and $Q$ \cite{Sultana2011,Fedele2002,Kourakis2005}. When $P$ and $Q$ are same sign ($P/Q>0$),
the evolution of the IAWs amplitude is modulationally unstable whereas when $P$ and $Q$ are
opposite sign ($P/Q<0$), the IAWs are modulationally stable in presence of the external perturbations.
The plot of $P/Q$ against $k$ yields stable and unstable domain for the IAWs.
The point, at which transition of $P/Q$ curve intersect with $k$-axis, is known as threshold
or critical wave number $k$ ($=k_c$). We have depicted $P/Q$ versus $k$ curve with $\mu_1$
for both fast and slow IA modes in Figs. \ref{2Fig:F1} and \ref{2Fig:F2}, respectively, and
it is obvious from these figures that (i) both stable and unstable domain can be found
for both fast and slow IA modes; (ii) the $k_c$ decreases (increases) with increasing in the values of
negative ion  mass for fast (slow) IA mode when $Z_-$, $Z_+$, and $m_+$ remain invariant.
So, the mass of the positive ion plays a vital role in recognizing the stabile domain of the IAWs.
\begin{figure}[!tbp]
  \centering
  \subfloat[]{\includegraphics[width=60mm]{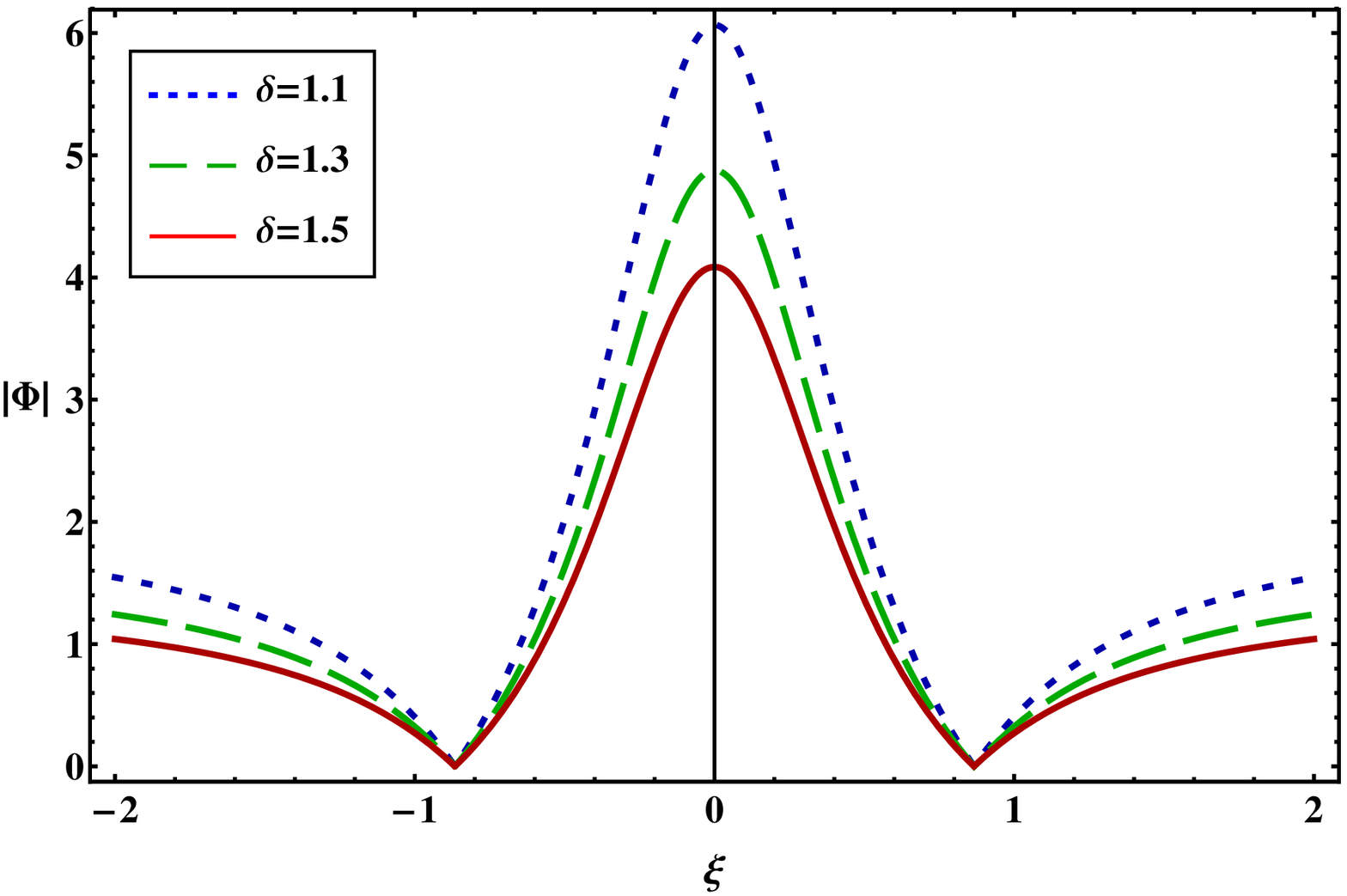}
  \label{2Fig:F5}}
   \hfill
  \subfloat[]{\includegraphics[width=60mm]{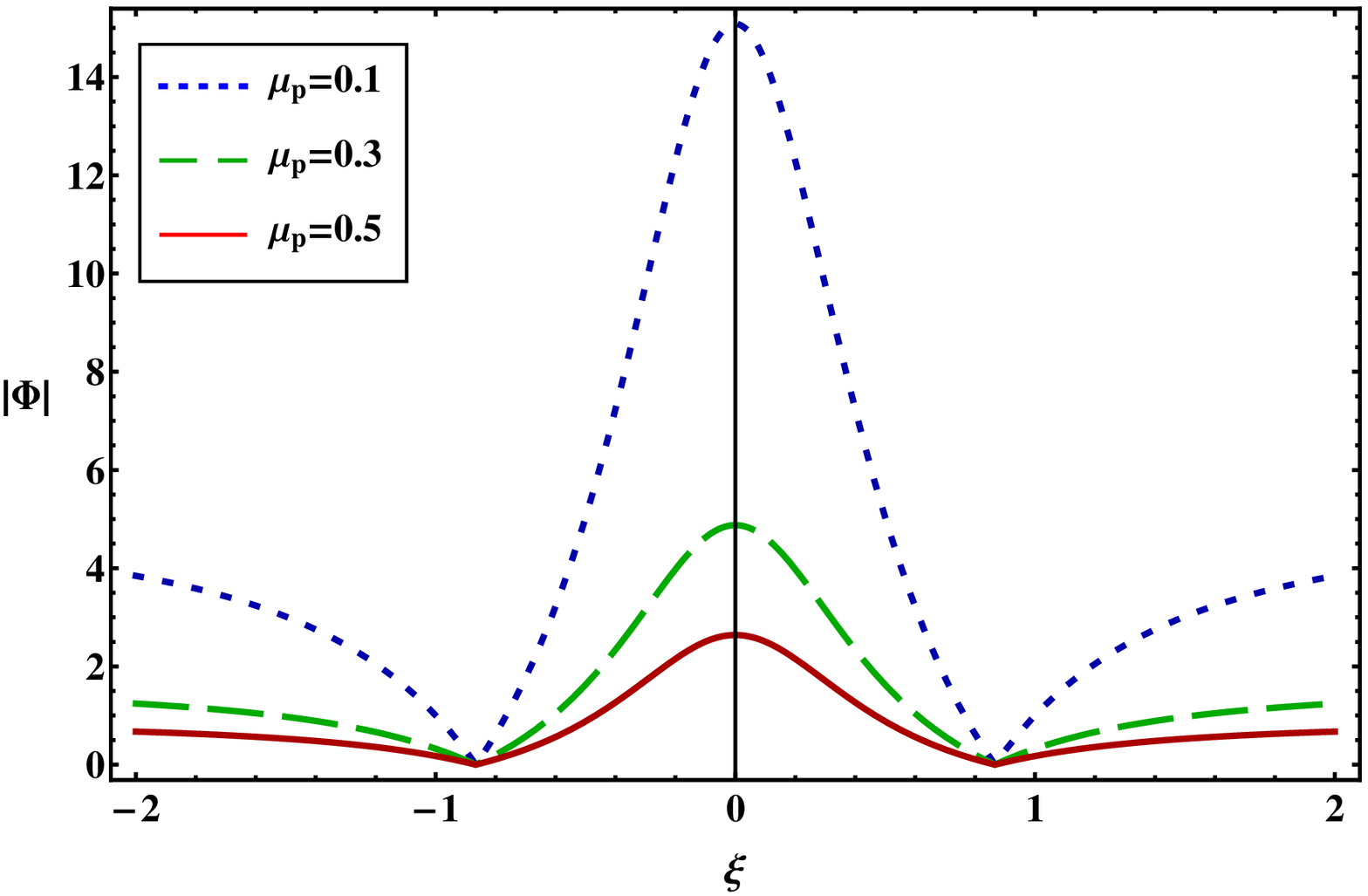}
  \label{2Fig:F6}}
  \caption{ The variation of $|\Phi|$ with $\xi$ at different values of (a) $\delta$ and $\mu_p=0.3$; (b) $\mu_p$ and $\delta=1.3$ ;
  along with $\mu_1=0.5$, $\mu_2=1.2$, $\sigma_1=0.005$, $\sigma_2=0.04$, $q=1.5$, $k=0.35$, $\tau=0$, and $\omega_f$.}
\end{figure}

The governing equation for the highly energetic IARWs in the unstable region ($P/Q>0$) can be written as \cite{Akhmediev2009,Ankiewiez2009}
\begin{eqnarray}
&&\hspace*{-1cm}\Phi(\xi,\tau)= \sqrt{\frac{2P}{Q}} \left[\frac{4(1+4iP\tau)}{1+16P^2\tau^2+4\xi^2}-1 \right]\mbox{exp}(2iP\tau).
\label{2eq:14}
\end{eqnarray}
We have numerically analyzed \eqref{2eq:14} in Figs. \ref{2Fig:F3}  and \ref{2Fig:F4}
to understand the effects of the mass and the number density of the negaitve/positive
ions, in fact the charge state of negaitve/positive ions, on the formation of the IARWs, and
it is clear from these figures that (a) the nonlinearity of the DPP medium  decreases,
by depicting smaller amplitude of the IARWs, with negative ion mass for constant values
of positive ion mass, charge state of the positive and negative ions (via $\mu_1$);
(b) the electrostatic IARWs potential  increases with the increase in the value of
negative ion number density ($n_{-0}$), but decreases with increase of the positive
ion number density ($n_{+0}$) for constant value of  $Z_+$ and $Z_-$ (via $\mu_2$);
(c) physically, negative ions lead to enhance the nonlinearity of the plasma medium
successively the amplitude of the electrostatic IARWs potential. This result agrees
with the result of El-Labany \textit{et al.} \cite{El-Labany2012} work.
\begin{figure}[!tbp]
  \centering
  \subfloat[]{\includegraphics[width=60mm]{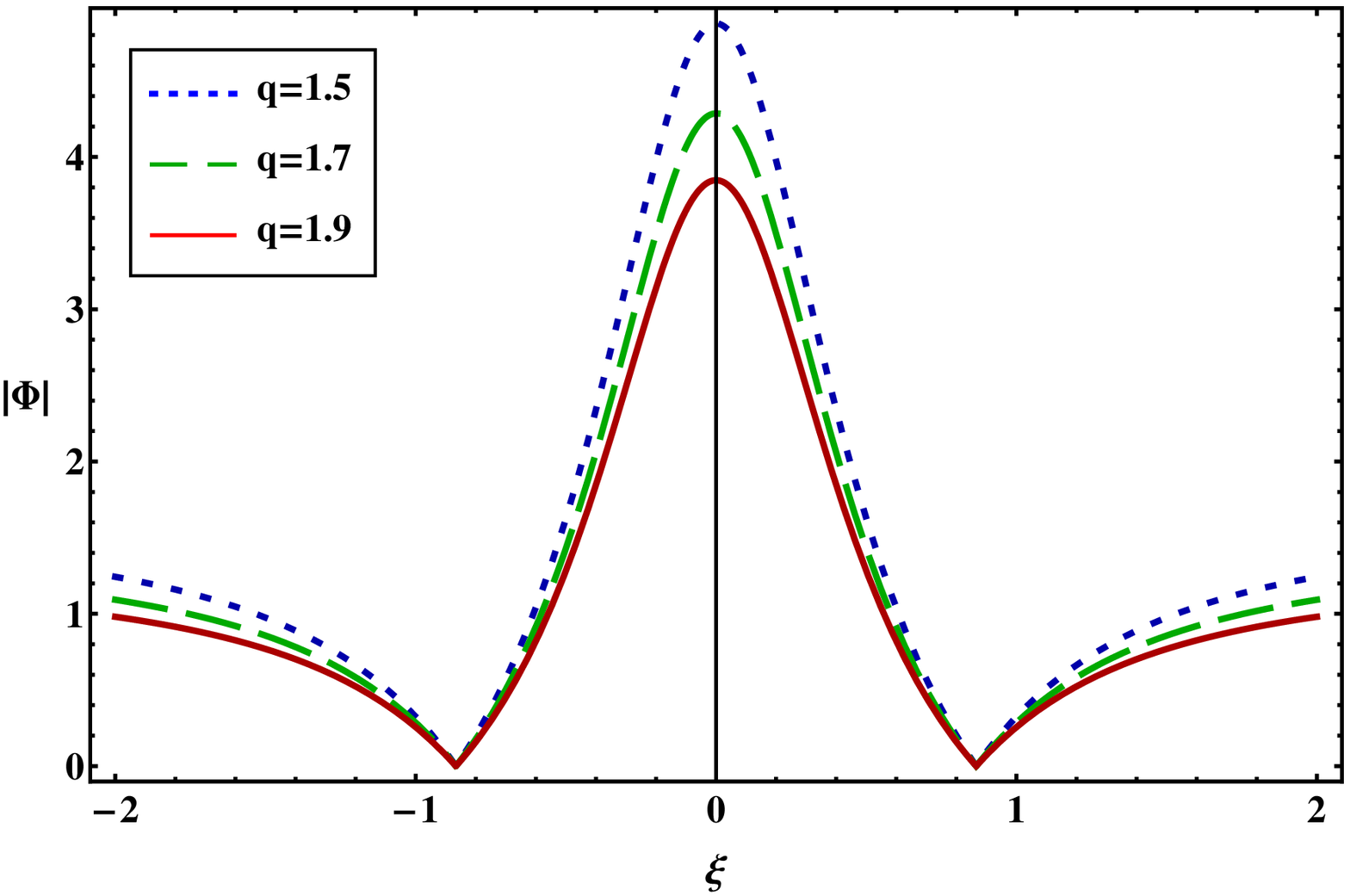}
  \label{2Fig:F7}}
   \hfill
  \subfloat[]{\includegraphics[width=60mm]{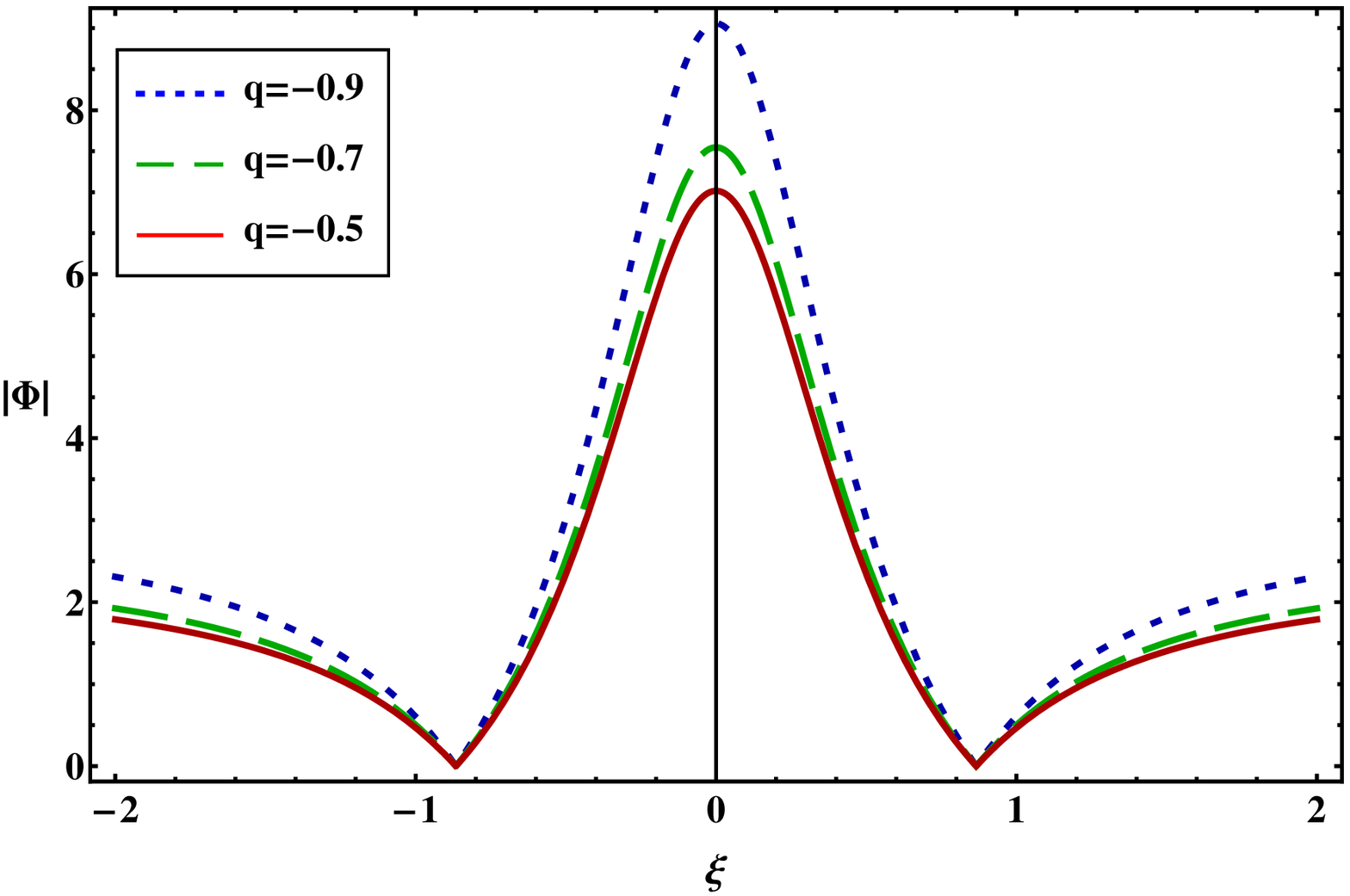}
  \label{2Fig:F8}}
  \caption{The variation of $|\Phi|$ with $\xi$ at different values of $q$ for (a) $q=+$ve; (b) $q=-$ve; along
  with $\delta=1.3$, $\mu_1=0.5$, $\mu_2=1.2$, $\mu_p=0.3$, $\sigma_1=0.005$, $\sigma_2=0.04$, $k=0.35$, $\tau=0$, and $\omega_f$.}
  \end{figure}

Figure \ref{2Fig:F5} is illustrated to demonstrate the effects of electron-to-positron temperature ratio on the IARWs.
It clearly shows that the electrostatic IARWs potential decreases with increasing electron-to-positron
temperature ratio (via $\delta$). Similarly, from  Fig. \ref{2Fig:F6} it can be shown that the electrostatic IARWs potential decreases with the increase in the value of positron number density ($n_{p0}$), but increases with increase of the negative ion number density ($n_{-0}$) for constant value of $Z_-$ (via $\mu_p$).

We have also numerically analyzed \eqref{2eq:14} in Figs. \ref{2Fig:F7} and \ref{2Fig:F8} to illustrate
the influence of $q-$distributed electrons on the formation of IARWs and it is obvious that
(i) the amplitude of the IARWs decreases with $q$ (for both $q>0$ and $q<0$); (ii) the amplitude
of electrostatic wave is independent of the sign of $q$, but dependent on the magnitude of $q$,
and this result is a good agreement with Chowdhury \textit{et al.} \cite{Chowdhury2017} work;
(iii) these figures, in fact, indicate a comparison between electrostatic potential for $q>0$ and $q<0$;
(iv) the amplitude of the IARWs electrostatic potential for same interval of positive $q$ is not same as negative $q$.
\section{Discussion}
\label{2sec:Discussion}
In this study, we have performed a nonlinear analysis of IAWs in a MPPs
consisting of inertialess $q$-distributed electrons and positrons, inertial adiabatic negative as well as positive ions.
The evolution of the IAWs is governed by the standard NLSE and the nonlinear $P$ and the dispersive $Q$ coefficients
represent the stable/unstable domain of the IAWs in presence of the external perturbation.
This theoretical investigation give rises to some noteworthy results that can be summed up as follows:
\begin{enumerate}
\item{Both fast and slow IA modes admit stable and unstable domains for the IAWs.}
\item{The nonlinearity of the MPPs  decreases with negative ion mass for constant
      values of positive ion mass, charge state of the positive and negative ions (via $\mu_1$).}
\item{The electrostatic IARWs potential  increases (decreases) with an increase in the value of
      negative ion number density (positive ion density) for invariant charge state of the positive and negative ions (via $\mu_2$).}
\item{IARWs potential decreases with increasing electron-to-positron temperature ratio (via $\delta$).}
\item{IARWs potential decreases (increases) with the increase in the value of positron number density
      (negative ion number density) for constant charge state of negative ion (via $\mu_p$)}.
\item{The amplitude and the width of the IARWs decrease with $q$ (for both $q>0$ and $q<0$).}
\end{enumerate}
Finally, we hope that the results from our present theoretical investigation may be helpful
in understanding the MI of IAWs and generation of the IARWs in laboratory plasmas, viz.
PI fullerene ($\rm C^+, C^-$) \cite{Sabry2008} or in the space plasma, viz. D-region ($\rm H^+, O_2^-$)
and F-region ($\rm H^+, H^-$) of the Earth's ionosphere \cite{Elwakil2010}, and Titan's atmosphere \cite{El-Labany2012}.
\section*{Acknowledgement}
S. Khondaker is grateful to the Bangladesh Ministry of Science and Technology for
awarding the National Science and Technology (NST) Fellowship.

\end{document}